\documentclass[fleqn,usenatbib]{mnras}

\usepackage{newtxtext,newtxmath}

\usepackage[T1]{fontenc}

\DeclareRobustCommand{\VAN}[3]{#2}
\let\VANthebibliography\thebibliography
\def\thebibliography{\DeclareRobustCommand{\VAN}[3]{##3}\VANthebibliography}

\usepackage{graphicx}	
\usepackage{amsmath}	

\newcommand{\Swift}{\textit{Swift}}

\usepackage{longtable}

\title[High energy mHz QPOs in 1A 0535+262]{High energy Millihertz quasi-periodic oscillations in 1A 0535+262 with Insight-HXMT challenge current models}

\author[Ma et al.]
{Ruican Ma,$^{1,2,3}$
Lian Tao,$^{1}\thanks{E-mail: taolian@ihep.ac.cn (IHEP)}$
Shuang-Nan Zhang,$^{1,2}$
Long Ji,$^{4}$
Liang Zhang,$^{1}$
Qingcui Bu,$^{5}$
Jinlu Qu,$^{1}$
\newauthor
Pablo Reig,$^{6,7}$
Mariano M\'endez,$^{3}$ 
Yanan Wang,$^{8}$  
Xiang Ma,$^{1}$ 
Yue Huang,$^{1}$ 
Mingyu Ge,$^{1}$ 
Liming Song,$^{1}$ 
\newauthor
Shu Zhang,$^{1}$ 
Hexin Liu,$^{1,2}$ 
Pengju Wang,$^{1,2}$ 
Lingda Kong,$^{1,2}$ 
Xiaoqin Ren,$^{1,2}$ 
Shujie Zhao,$^{1,2}$ 
Wei Yu,$^{1,2}$
\newauthor
Zixu Yang,$^{1,2}$ 
Panping Li,$^{1,2}$ 
Shumei Jia$^{1}$ 
\\
$^{1}$Key Laboratory of Particle Astrophysics, Institute of High Energy Physics, Chinese Academy of Sciences, Beijing 100049, China\\
$^{2}$University of Chinese Academy of Sciences, Chinese Academy of Sciences, Beijing 100049, China\\
$^{3}$Kapteyn Astronomical Institute, University of Groningen, P.O. BOX 800, 9700 AV Groningen, The Netherlands\\
$^{4}$School of Physics and Astronomy, Sun Yat-Sen University, Zhuhai, 519082, China\\
$^{5}$Institut f\"ur Astronomie und Astrophysik, Kepler Center for Astro and Particle Physics, Eberhard Karls Universit\"at, 72076 T\"ubingen, Germany\\
$^{6}$Institute of Astrophysics, Foundation for Research and Technology-Hellas, 71110 Heraklion, Crete, Greece\\
$^{7}$University of Crete, Physics Department, 70013 Heraklion, Crete, Greece\\
$^{8}$Physics and Astronomy, University of Southampton, Southampton, Hampshire SO17 1BJ, UK
}

\date{Accepted 2022 September 20. Received 2022 September 5; in original form 2022 August 5}

\pubyear{2022}

\begin{document}
\label{firstpage}
\pagerange{\pageref{firstpage}--\pageref{lastpage}}
\maketitle

\begin{abstract}
We studied the millihertz quasi-periodic oscillation (mHz QPO) in the 2020 outburst of the Be/X-ray binary 1A 0535+262 using \textit{Insight}-HXMT data over a broad energy band. The mHz QPO is detected in the 27$-$120\,keV energy band. The QPO centroid frequency is correlated with the source flux, and evolves in the $35-95$\,mHz range during the outburst. The QPO is most significant in the 50$-$65\,keV band, with a significance of $\sim{8\sigma}$, but is hardly detectable ($<2\sigma$) in the lowest (1$-$27\,keV) and highest ($>120$\,keV) energy bands. Notably, the detection of mHz QPO above 80\,keV is the highest energy at which mHz QPOs have been detected so far. The fractional rms of the mHz QPO first increases and then decreases with energy, reaching the maximum amplitude at 50$-$65\,keV. In addition, at the peak of the outburst, the mHz QPO shows a double-peak structure, with the difference between the two peaks being constant at $\sim$\,0.02\,Hz, twice the spin frequency of the neutron star in this system. We discuss different scenarios explaining the generation of the mHz QPO, including the beat frequency model, the Keplerian frequency model, the model of two jets in opposite directions, and the precession of the neutron star, but find that none of them can explain the origin of the QPO well. We conclude that the variability of non-thermal radiation may account for the mHz QPO, but further theoretical studies are needed to reveal the physical mechanism.
\end{abstract}




\section{Introduction}
\label{sec:intro}

Neutron-star X-ray binaries (NSXRBs) are systems composed of a neutron star and a donor star. When the mass of the donor star is greater than $\sim8~M_{\odot}$, NSXRBs are classified as high-mass systems. According to the class of the companion star, high-mass systems can be further divided as Be/XRBs (BeXBs) and supergiant XRBs \citep[SGXBs; e.g.][]{Reig2011}. Different from SGXBs, BeXBs are usually transient sources and can undergo two types of X-ray outbursts, namely type I (normal) and type II (giant) outbursts. Type I outbursts are generally periodic or quasi-periodic, and have a short duration of $\sim$0.2$-$0.3 orbital period. The flux of the outburst peaks near periastron, reaching a luminosity of less than $10^{37}\rm{\,erg\,s^{-1}}$. Type II outbursts are characterized by higher X-ray luminosity (typically greater than $10^{37}\rm{\,erg\,s^{-1}}$) and longer duration (usually several orbits) than type I outbursts. Moreover, their occurrence does not depend on the orbital phase \citep{Priedhorsky1983, Reig2011}.

During giant outbursts, the compact star in BeXBs exhibits some interesting spectral (e.g. cyclotron line) and timing (e.g. spin evolution) features. In addition to periodic behavior, the aperiodic variability, especially millihertz Quasi-Periodic Oscillations (mHz QPOs: 1$-$400\,mHz; \citealt{Boroson2000, Mukerjee2001, Bozzo2009}), are a common feature that help us understand the accretion process. MHz QPOs are more commonly found in transient rather than persistent sources \citep{Dugair2013}. Moreover, mHz QPOs are also transient in nature \citep{Dugair2013}. For some sources, such as 4U 1626--67 \citep{Kaur2008} and 1A 1118--61 \citep{Devasia2011a}, the mHz QPO can be detected during most of the outburst, while for some other sources, such as Cen X--3 \citep{Raichur2008}, KS 1947+300 \citep{James2010} and 4U 1901+03 \citep{James2011}, these QPOs are only found during part of the outburst. 

Although various models have been proposed to explain the origin of mHz QPOs, we still do not fully understand its nature. The most common models are the beat-frequency (BF) model and the Keplerian-frequency (KF) model. In the BF model, blobs of matter from inhomogeneities in the disk orbit at the Keplerian frequency ($\nu_{\rm k}$) at the inner boundary of the accretion disk, and are accreted by the NS at a rate that is modulated by the rotating magnetic field, thus the QPO frequency ($\nu_{\rm{q}}$) is derived from the beat between $\nu_{\rm{k}}$ and the NS spin frequency ($\nu_{\rm{s}}$), i.e. $\nu_{\rm{q}}=\nu_{\rm{k}}-\nu_{\rm{s}}$ \citep{Alpar1985, Shaham1987}. For the KF model, the mHz QPO comes from the obscuration of the X-ray flux by structures in the accretion disk, and thus $\nu_{\rm{q}}$ is equal to $\nu_{\rm{k}}$ at the inner accretion disk \citep{vdKlis1987}. 

One key factor that limits our understanding of the mHz QPO is that so far X-ray observatories have not been very sensitive to photons above $\sim 30-40$\,keV and thus the QPO properties over a broad energy range, especially at high energies, are poorly understood. There are only a few detailed reports on the evolution properties of QPO with energy. At present, the evolution of the fractional rms amplitude of the QPO with energy has only been reported for six sources, KS 1947+300 \citep{James2010}, GX 304$-$1 \citep{Devasia2011b}, Cen X$-$3 \citep{Raichur2008}, 1A 1118$-$61 \citep{Devasia2011a}, XTE J1858+034 \citep{Mukherjee2006} and V0332+53 \citep{Qu2005}, but the observation of these sources are all in a relatively narrow energy band, below 40\,keV. \par

Compared with the previous generation of observatories, \textit{Insight}-HXMT is a mission with a broad energy coverage (1--250\,keV) and has the largest effective area above 30\,keV \citep{Zhang2020}. Therefore, we expect that using the \textit{Insight}-HXMT data the properties of mHz QPOs can be further studied. A good opportunity came up in 2020, as the accreting pulsar 1A 0535+262 underwent a giant outburst \citep{Mandal2020,Pal2020}. \par

1A 0535+262 is a transient BeXB pulsar discovered in 1975 with \textit{Ariel V} \citep{Rosenberg1975,Coe1975}. The orbital period of the binary system and pulse period of the NS are $\sim 111$\,days and $\sim 104$\,s, respectively \citep{Finger1996}. The distance of the source is 2.13\,kpc \citep{Bailer2018}. Since its discovery, 1A 0535+262 underwent at least 10 giant outbursts \citep{Camero2012}. Its 2020 outburst has so far the brightest one, with a flux as high as 11 Crab in the \Swift/BAT light curves. In this outburst, multiwavlength monitoring campaigns in radio, optical and X-ray have been conducted \citep{vdEijnden2020,Arabaci2020,Jaisawal2020}. In particular, \textit{Insight}-HXMT has performed 1910\,ks high statistics observations, which covered the entire outburst in the 1$-$250\,keV energy band. 

The properties of the mHz QPO in 1A 0535+262 in the two giant outbursts of 1994 and 2009 were reported by \citet{Finger1996} and \citet{Camero2012}, respectively. The mHz QPO was detected most of the time during those outbursts, with the frequency varying from 30\,mHz to 70\,mHz \citep{Finger1996,Camero2012}. Moreover, the QPO frequency was proportional to the spin-up rate and the pulsed flux \citep{Finger1996,Camero2012}, which is consistent with the expectation of the BF and KF models \citep{Finger1996}. However, further research by \citet{Camero2012} showed that the mHz QPO became stronger with increasing energy but disappeared below 25\,keV. Such an energy-dependent behavior questions both models, yet more studies are needed in a broader energy band. We thus expect that the bright 2020 outburst, with the broad and high statistics \textit{Insight}-HXMT observations, can procide a deeper understanding of the mHz QPO in this source.

In this work, we report the properties of the mHz QPO in 1A 0535+262 during the 2020 outburst. We studied the $1-250$\,keV power spectra and detected the mHz QPO above 80\,keV, which is highest energy at which the mHz QPO have been detected among X-ray pulsars. We investigated the energy-dependent properties of the mHz QPO and their evolution over the outburst to have a better understanding of the mechanism that produces the QPO. The paper is structured as follows. The observations and data reduction are described in Section~\ref{sec:obs}. The results are presented in Section~\ref{sec:res}, discussed in Section~\ref{sec:dis}, and summarized in Section~\ref{sec:con}.

\section{Observations and Data analysis}
\label{sec:obs}

\textit{Insight}-HXMT is China's first X-ray astronomy telescope, carrying the low-energy X-ray telescope \citep[LE, 1$-$10\,keV, 384\,cm$^2$;][]{Chen2020}, the medium-energy X-ray telescope \citep[ME, 8$-$35\,keV, 952\,cm$^2$;][]{Cao2020} and the high-energy X-ray telescope \citep[HE, 27$-$250\,keV, 5100\,cm$^2$;][]{Liu2020}. 1A 0535+262 was monitored by \textit{Insight}-HXMT from November 6 (MJD 59159) to December 24, 2020 (MJD 59207). In this work, we mainly analyze the data from the core program proposal P0314316, because the statistics of the core observations are much better than the guest observations (e.g., the exposure time per observation was 90$-$190\,ks versus 10\,ks). The core program observations are from MJD 59167 to 59207, with a total exposure of 1910\,ks (see Table~\ref{tab:obs_info} for details), covering part of the rise, the outburst peak and the full decay phase.

We used the \textit{Insight}-HXMT Data Analysis Software (HXMTDAS) v2.04 for the data reduction. The good time intervals (GTIs) are selected as follow: (1) Earth elevation angle $>10^{\circ}$; (2) geomagnetic cutoff rigidity $>8$\,GV; (3) pointing offset angle $<0.04^{\circ}$; (4) at least 300\,s away from the South Atlantic Anomaly (SAA). The background of the HE, ME and LE instruments are estimated using the tool HEBKGMAP, MEBKGMAP and LEBKGMAP, respectively, based on the \textit{Insight}-HXMT background model \citep{Liao2020a, Guo2020, Liao2020b}. Although the data of the ME and LE instruments are also analyzed, we mainly report the results of the HE instrument in this paper, since the mHz QPO is not significantly detected by the LE and ME telescopes (with a significance generally below 2$\sigma$, see Section~\ref{sec:QPO_energy} for more details). Crab (R.A.=05$^{\rm h}$34$^{\rm m}$31.$^{\rm s}$94, Decl.=$22^{\circ}$00$^{'}$52.$^{''}$2; J2000) is close to 1A 0535+262 (R.A.=05$^{\rm h}$38$^{\rm m}$54.$^{\rm s}$57, Decl.=$26^{\circ}$18$^{'}$56.$^{''}$8; J2000), but it does not affect the timing analysis, given that it only accounts for $\sim 2.4$\% of the detected photons.

In order to improve the statistical significance, the HE data of one day (usually about 5$-$8 exposure IDs\footnote{To reduce the single file size, an observation is artificially split into multiple segments (named exposure), which is only a time segmentation in an observation without the traditional exposure concept of camera or CCD image instrument.} per day) are merged. We then divide the merged HE data into eight energy bands, namely $27-35$\,keV, $35-40$\,keV, $40-50$\,keV, $50-65$\,keV, $65-80$\,keV, $80-120$\,keV and $120-250$\,keV. The power-density spectra (PDS) of each dataset is generated using the {\tt POWSPEC} tool in HEASOFT version 6.27. Since the QPO signal is found around 0.1\,Hz, we produce the PDS in the range of $0.004-0.5$\,Hz. The PDS are normalized to the square mean intensity \citetext{\citealp[]{Miyamoto1991}, \citealp[see also][]{Belloni1990}}. The dead time ($\tau$) correction is not considered because $\tau$ of HE is $\sim 20\,\rm{\mu s}$ and the frequency range of the PDS is much lower than 1/$\tau$. The pulsation and its harmonics would pollute the PDS, we therefore remove the frequency bins that are affected by these components. We also compare the PDS results with the method adopted in \citet{Reig2022}, in which the pulse flux is subtracted before generating the PDS, and found that the results are consistent. 

The PDS are fitted in {\sc xspec}  \citep[version 12.11.1;][]{Arnaud1996}. A narrow Lorentzian is used to fit the QPO, if found, while two zero-centered Lorentzians and a power-law component with constant are adopted to describe, respectively, the low-frequency and the Poisson noise \citep{Nowak2000,Belloni2002}. After considering the background contribution, the fractional rms of the QPO is calculated as $\sqrt{N}*(I_{\rm S}+I_{\rm B})/I_{\rm S}$ over the frequency range $\nu\pm{\rm{HWHM}}$, where $I_{\rm S}$ is the source count rate, $I_{\rm B}$ is the background count rates, $N$ is the power normalized, $\nu$ is the QPO centroid frequency and HWHM is its half width at half maximum.

\section{Results}
\label{sec:res}

\subsection{Energy-dependent behaviour of 1A 0535+262}
\label{sec:QPO_energy}

\begin{figure}
\centering
\includegraphics[width=\columnwidth]{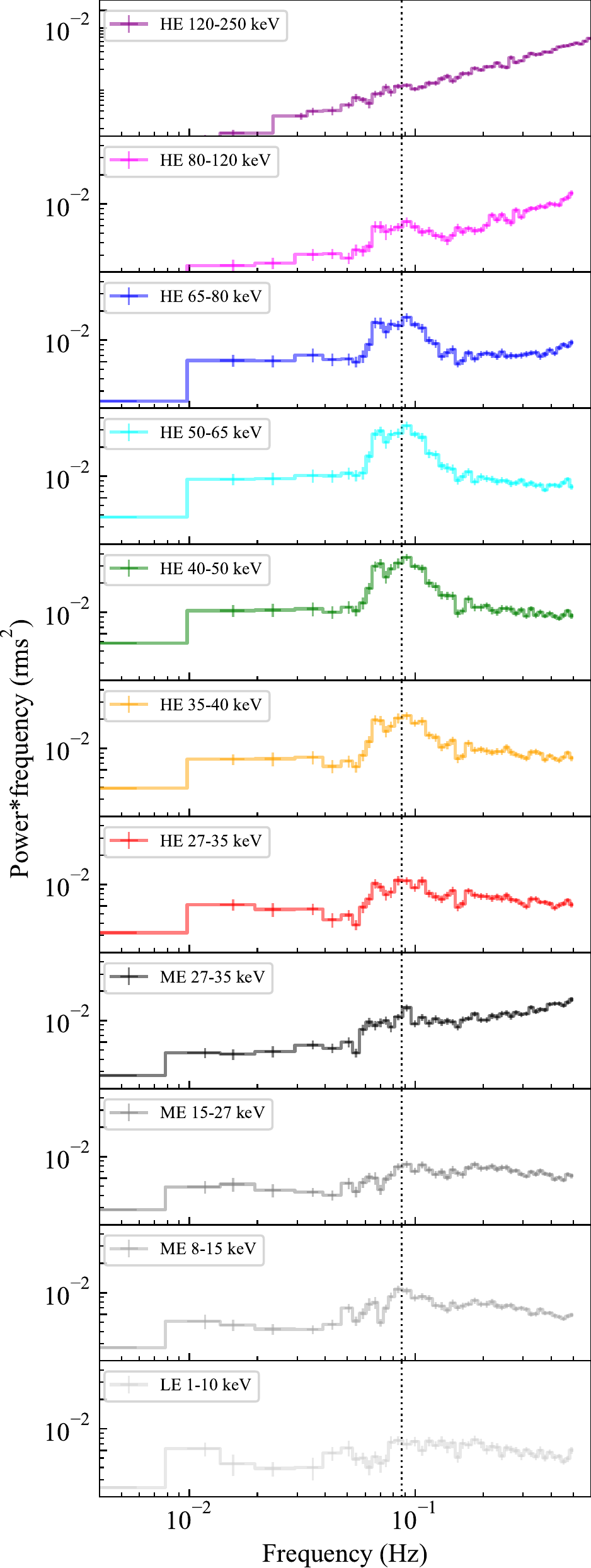} \\
\caption{PDS of 1A 0535+262 in different energy bands on MJD 59169. The vertical dashed line represents the mHz QPO centroid frequency.
 \label{fig:PDS}}
\end{figure}

\begin{figure}
\centering
\includegraphics[width=\columnwidth]{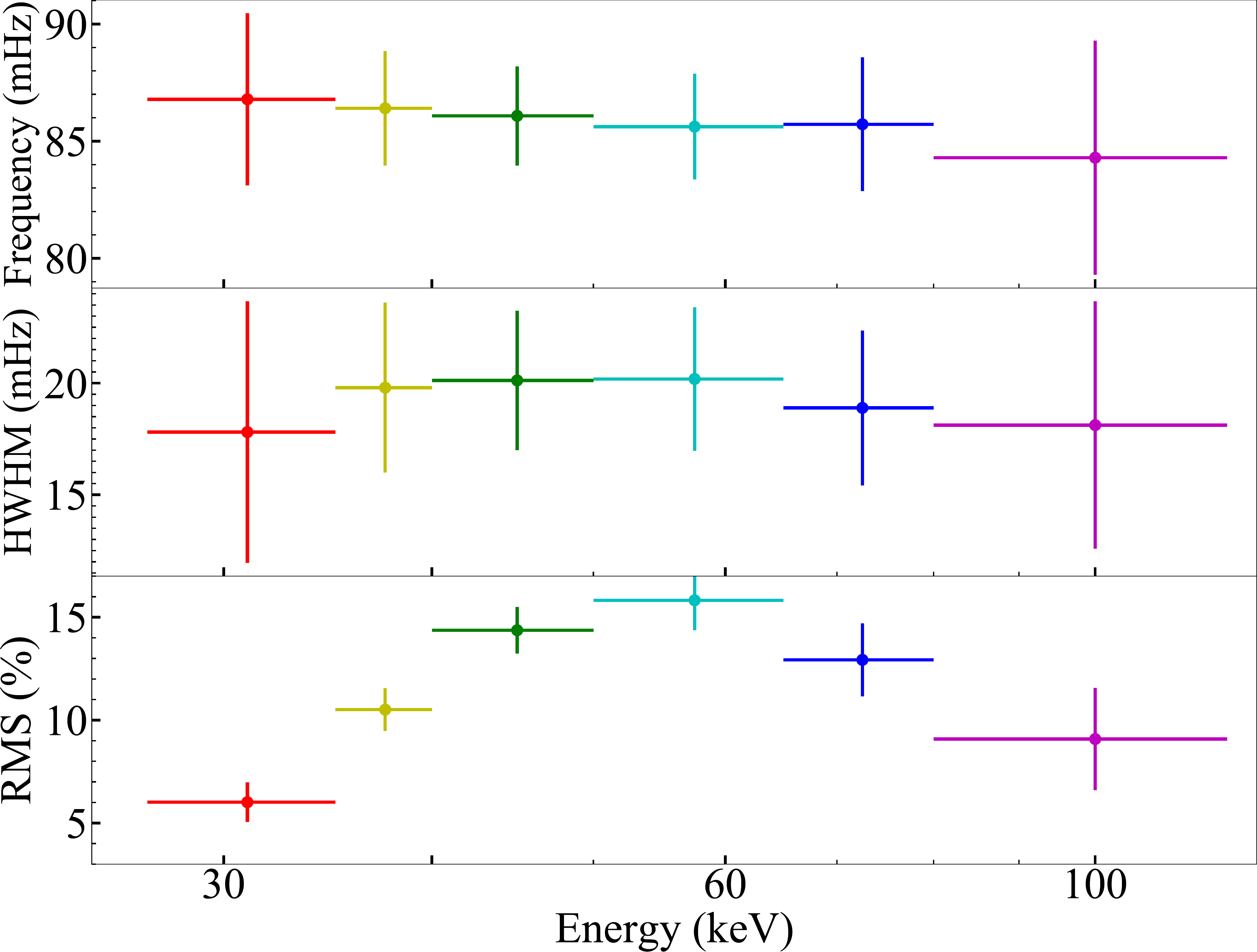} \\ %
\caption{Centroid frequency, HWHM and fractional rms amplitude of the mHz QPO of 1A 0535+262 in different energy bands. Different colors represent different energy bands, same as Figure~\ref{fig:PDS}. 
\label{fig:QPO_energy_evolution}}
\end{figure}

\begin{figure}
\centering
\includegraphics[width=\columnwidth]{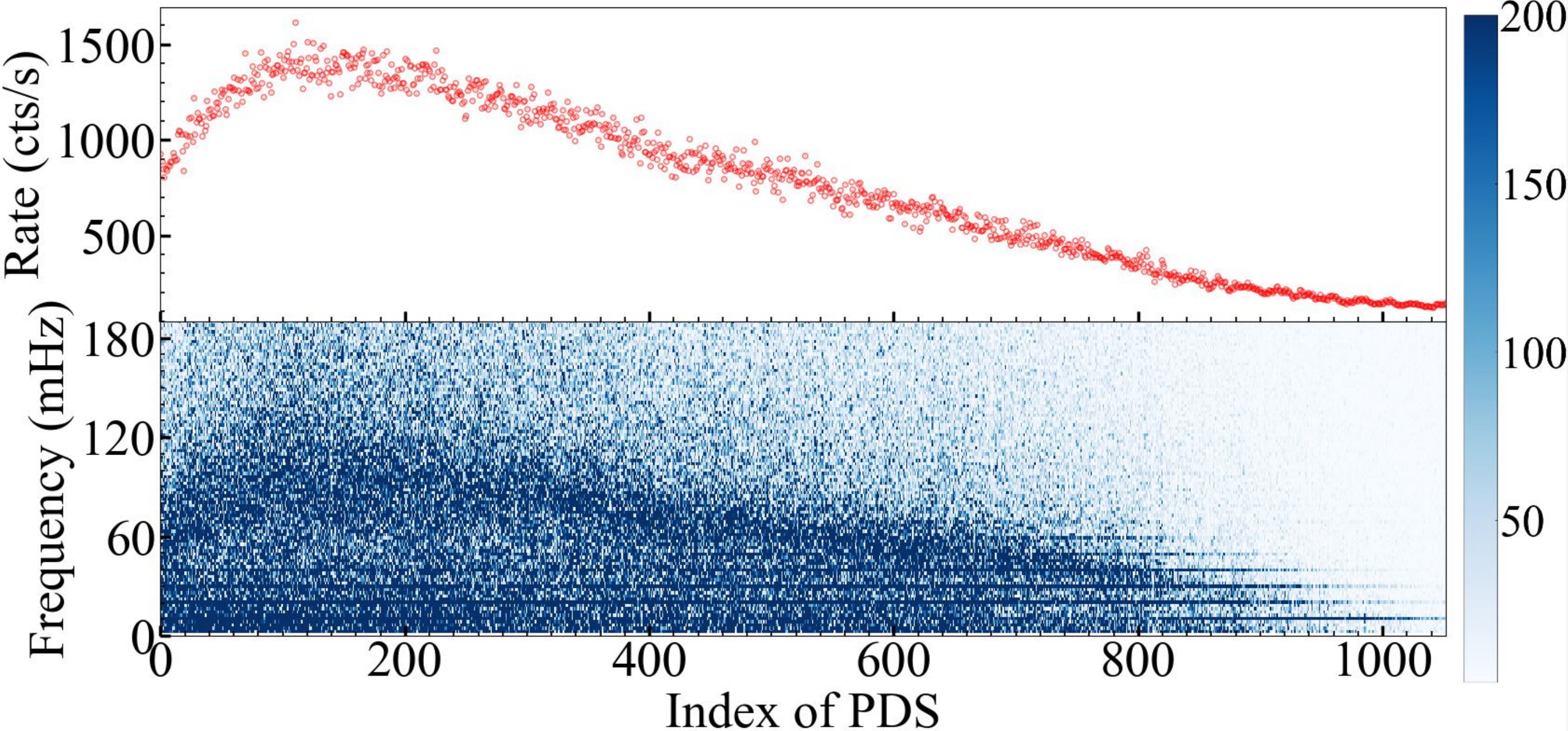} \\
\caption{Light curve and dynamic PDS of 1A 0535+262 in the 40$-$65\,keV band from MJD 59167 to 59207. In the light curve, the bin size is 512\,s. The darker the color in the dynamic PDS, the greater the power value. The horizontal lines in the bottom panel represent the spin and harmonics of the pulsar.
\label{fig:DyPDS}}
\end{figure}

\begin{figure}
\centering
\includegraphics[width=\columnwidth]{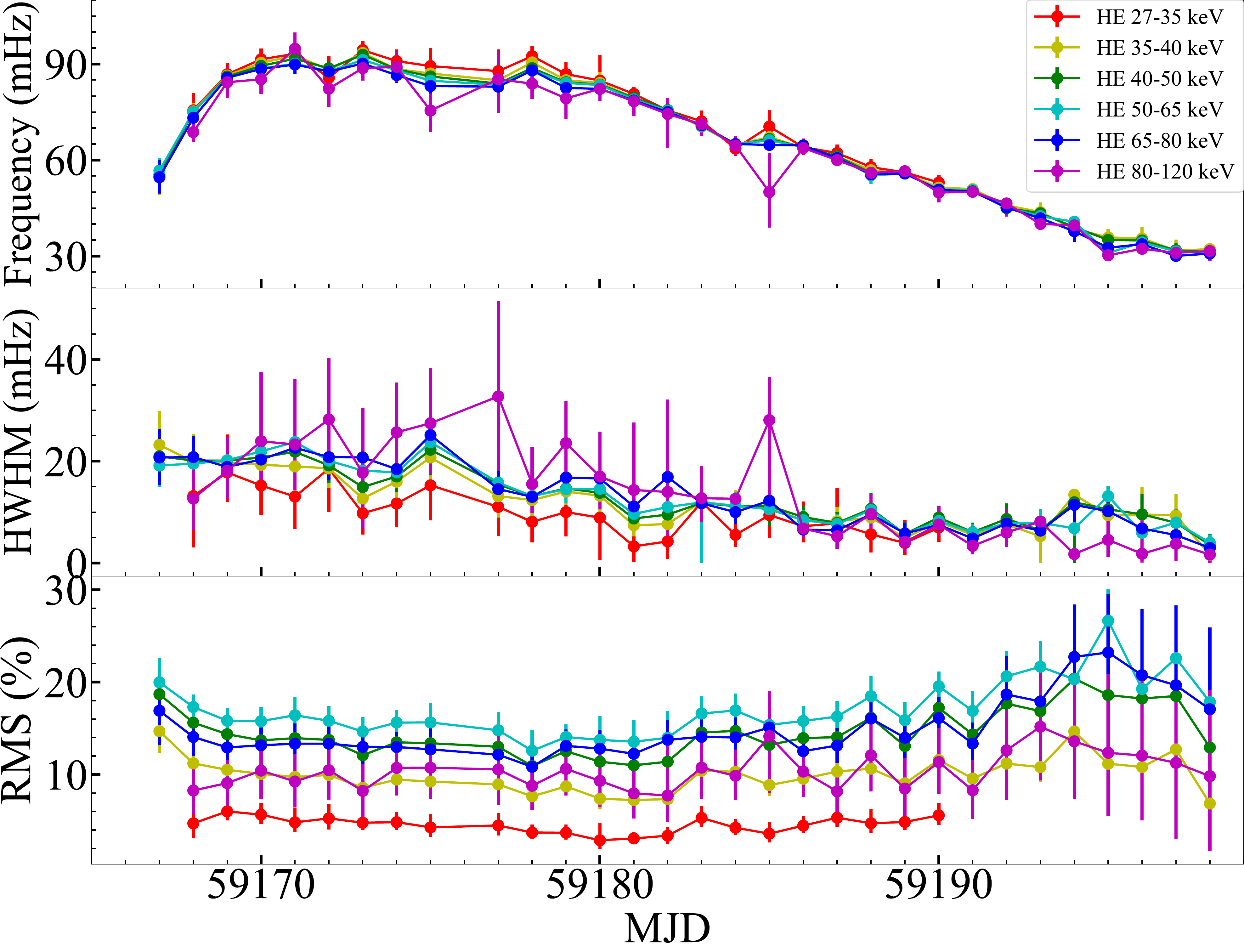} \\ %
\caption{Evolution of the centroid frequency, HWHM and fractional rms amplitude of the mHz QPO in 1A 0535+262 during the 2020 outburst. Different colors represent different energy bands, same as in Figure~\ref{fig:PDS}.
\label{fig:QPO_time_evolution}}
\end{figure}

Similar to the 1996 and 2009 outbursts \citep{Finger1996,Camero2012}, a QPO with centroid frequency 30$-$95\,mHz also appears in the 2020 outburst. The typical PDS (on MJD 59169) in different energy bands are shown in Figure~\ref{fig:PDS}. With data from \textit{Insight}-HXMT, we present the QPO behavior in a broader band and find that the QPO is strongest in the $40-50$\,keV and $50-65$\,keV energy bands, in which the significance can be as high as 7$\sigma$ and 8$\sigma$, respectively. In the lower and higher energy bands, the QPO is less significant, but we still can detect the QPO in the $80-120$\,keV energy band with a significance larger than 3$\sigma$. Notably, this is the highest energy band at which a mHz QPO has been so far detected. We do not detected significant QPO signal in the LE ($1-10$\,keV) and ME ($8-35$\,keV) telescopes, and also in HE above 120\,keV. Even in the overlapping energy band (27$-$35\,keV) between ME and HE, the QPO is not significant in the ME (less than 2$\sigma$), due to the effective area of ME being just a tenth of HE.  

Different from the non-detection of the QPO in the low energy band (see Section~\ref{dis:QPO_energy} for more details), the QPO not being detected in the 120$-$250\,keV band may be due to the lower count rate of the source in the high-energy band. Using Equation (2.17) in \citet[]{Lewin1988} \citep[see also, ][]{vdKlis1989}, assuming that the QPO has the same rms amplitude and FWHM in the 120$-$250\,keV band as in the 80$-$120\,keV band, to detect it we would have needed a count rate in the highest band that is $\sim$15 times higher than what we observe, therefore we cannot exclude that the QPO is also present in the highest energy band in our observation.

As shown in the top two panels of Figure~\ref{fig:QPO_energy_evolution}, the centroid frequency and HWHM of the QPO are nearly constant in different bands. The fractional rms amplitude as a function energy is shown in the third panel of Figure~\ref{fig:QPO_energy_evolution}. The rms amplitude first increases from 6\% to 16\% and then decreases to 9\% with increasing energy, with a maximum in the $50-65$\,keV band, showing an arch-shaped trend. It should be noticed that, in order to keep the consistency of the fitting model throughout the outburst, we use the fitting parameters of the QPO with only one Lorentzian even at the peak outburst, unless otherwise specified (see Section~\ref{sec:double_peak} for more details).

\subsection{Outburst evolution}

The light curve and dynamic PDS of the $40-65$\,keV band are shown in Figure~\ref{fig:DyPDS}. The QPO is visible throughout the outburst, until the source count rate is too low (below 60 cts\,$\rm{s^{-1}}$ in the 50$-$65\,keV energy band). As the outburst evolves, the QPO centroid frequency first increases rapidly from $\sim$ 55\,mHz to $\sim$ 95\,mHz, and then gradually decreases to $\sim$ 30\,mHz, following the evolution trend of the light curve (see Figure~\ref{fig:QPO_time_evolution}). As the 2020 outburst is the brightest one so far detected of 1A 0535+262, the QPO frequency range is broader than that of the 1994 and 2009 outbursts (30--70\,mHz) \citep{Finger1996,Camero2012}. Moreover, as shown in the lower panel of Figure~\ref{fig:DyPDS}, the spin fundamental and harmonics do not affect the power around the QPO during the outburst peak, and only weakly contribute to the power at the end of outburst.

The evolution of the QPO in different energy bands is similar (see Figure~\ref{fig:QPO_time_evolution}), varying from 30 to 95\,mHz. The evolution of the QPO HWHM and fractional rms amplitude are given in Figure~\ref{fig:QPO_time_evolution}. During the outburst, the HWHM decreases gradually from $\sim$ 24\,mHz to $\sim$ 4\,mHz. The fractional rms first showed a downward trend, and then gradually increased (although with large error bars).

\begin{figure*}
\centering
\includegraphics[width=\textwidth]{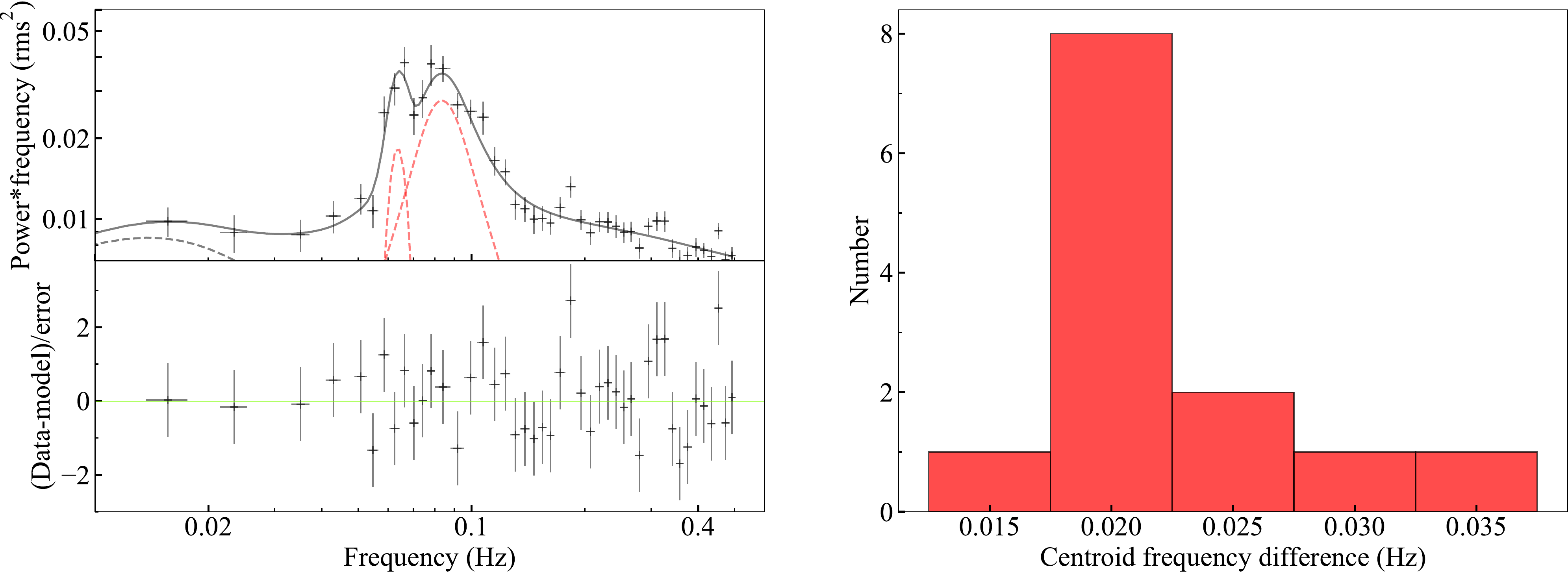} \\
\caption{Left panel: double-peak structure in a typical PDS (MJD 59168) of 1A 0535+262. The broad Lorentzian component and the narrow gaussian component are marked in red. Right panel: distribution of the centroid frequency difference between the two peaks in the PDS.
\label{fig:double peak}}
\end{figure*}

\subsection{Double-peak structure of the QPO}
\label{sec:double_peak}

During the peak of the outburst (MJD 59167--59176), a double-peak structure appears around the QPO frequency in the PDS (see the left panel of Figure~\ref{fig:double peak}). The existence of the double-peak structure does not depend on the selection of parameters when generating PDS, i.e., the time resolution, geometrical series rebin factor and duration of the time segment used to generate PDS. By dividing the light curve during the double-peak period into short segments and generating dynamical power spectra, we also exclude the possibility that the structure comes from the fast evolution of the QPO frequency when the source rate changes. 

There are 4 observation IDs (P0314316001$-$P0314316004) during the peak of outburst, including 59 exposure IDs (P031431600101$-$P031431600414; see table~\ref{tab:obs_info} for more details). In order to increase the data signal-to-noise ratio, we merged the $40-65$\,keV data of every 4 exposure IDs and obtained 15 datasets, and find that the double-peak structure appears in 13 datasets among them. For these datasets with double-peak structure, the PDS are fitted with a relatively broader Lorentzian (width $\sim$ 0.02\,Hz) plus a narrower Gaussian (width $\sim$ 0.003\,Hz). In some datasets, if the structure is weak, we fix the width of the broader Lorentzian at 0.02\,Hz. As shown in Figure~\ref{fig:double peak}, the two-component model fits the QPO well. Moreover, we note that the difference in the centroid frequency between the two peaks is always around 0.02\,Hz (see the right panel of Figure~\ref{fig:double peak}), which is twice the spin frequency of the pulsar.

\section{Discussion}
\label{sec:dis}

With the high statistic and high cadence observations provided by \textit{Insight}-HXMT, a QPO at 30$-$95\,mHz has been detected during the 2020 giant outburst of 1A 0535+262. Notably, we detect the mHz QPO above 80\,keV for the first time, which is highest energy mHz QPO ever found in any of the source. We studied the QPO properties with unprecedented energy coverage (1$-$250\,keV). Below, we discuss the origin of the mHz QPO based on the observational facts, especially the energy-dependent behavior.

\begin{figure}
\centering
\includegraphics[width=\columnwidth]{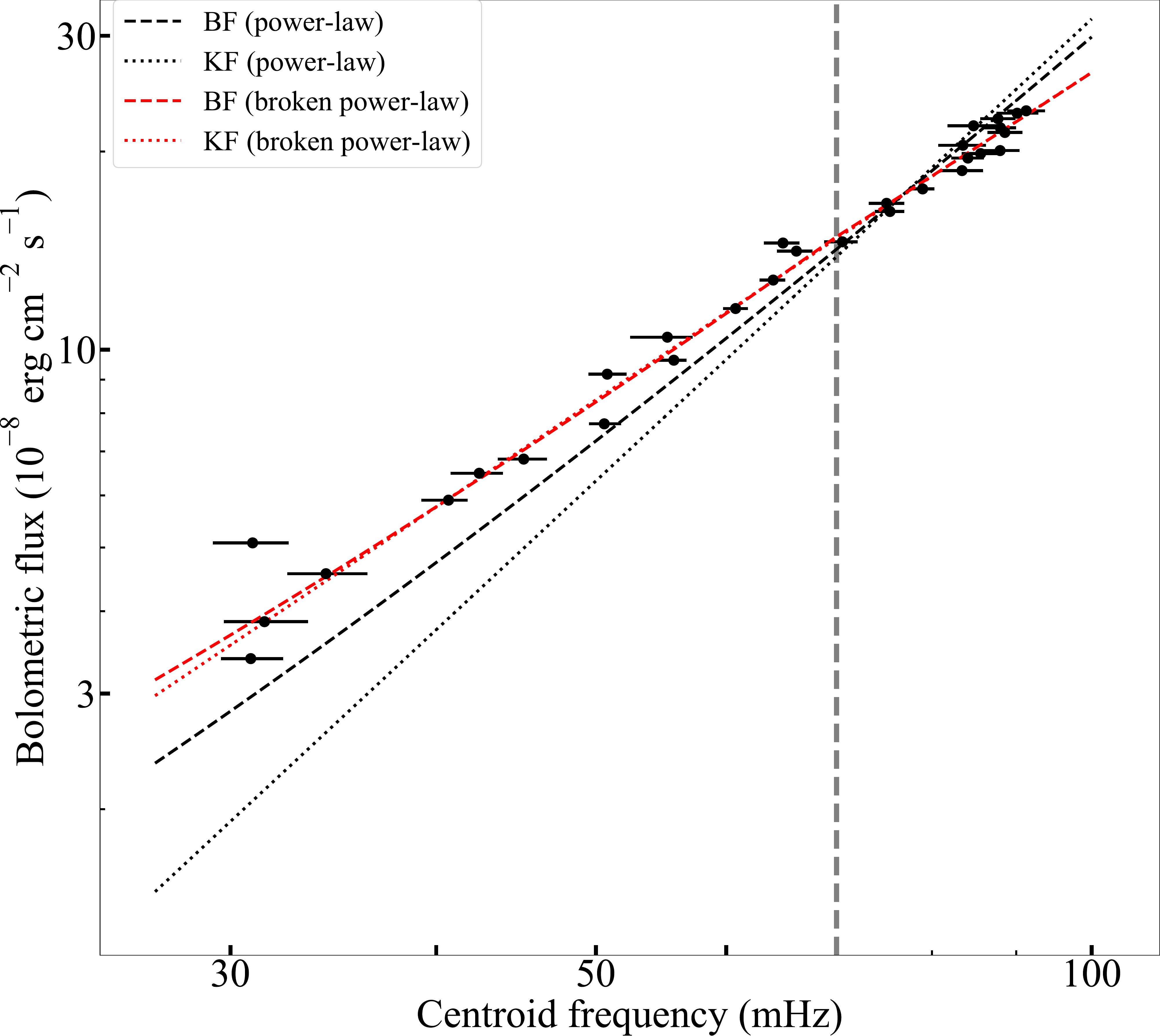} \\ 
\caption{Relation between the QPO centroid frequency and the bolometric flux of 1A 0535+262. The black dashed and dotted lines represent the predicted relation given by Equation~(\ref{equ1}), for the BF and KF models, respectively. The red dashed and dotted lines correspond to the broken power-law for the BF and KF models, respectively. The grey vertical line represents $\nu_{\rm k}=70$\,mHz.
\label{fig:QPO_flux}}
\end{figure}

\subsection{BF and KF models}
\subsubsection{Relationship between flux and QPO frequency}
\label{dis:QPO_frequency}

The BF and KF models are the two leading models to explain the mHz QPOs in accretion-powered X-ray pulsars. The BF model has successfully explained the origin of the QPOs in 4U 1901+03 \citep{James2011}, XTE J1858+034 \citep{Paul1998,Mukherjee2006} and EXO 2030+375 \citep{Angelini1989}, while the KF model could explain the origin of the mHz QPO in SAX J2103.5+4545 \citep{Inam2004} and some other sources. From the positive correlation between the source flux and the QPO frequency, \citet{Finger1996} concluded that both the BF and KF models can explain the generation of the mHz QPO in 1A 0535+262. The 2020 outburst of this source is the brightest one so far, thus it is essential to review this relationship over a larger range of flux and QPO frequency. 

The relationship between flux and Keplerian frequency is given as follows \citep{Finger1996},
\begin{equation}
\label{equ1}
    F= {\aleph}_{F}\nu^{7/3}_{k}\,{\rm with}~\aleph_{F}=1.4\beta d^{-2}K^{7/2}B^{2}R^{5}(GM)^{-2/3},
\end{equation}
where $F$ is the bolometric flux, $\nu_{\rm{k}}$ is the Keplerian frequency, $\beta$ accounts for the beaming effect, $d$ is the source distance, $B$ is the magnetic filed, $R$ is the NS radius, $G$ is the gravitational constant and $M$ is the NS mass. $K=(GM)^{10/21}(4\pi^{2}\nu^{2}_{\rm k})^{-1/3}\mu^{-4/7}\Dot{M}^{2/7}$ is the ratio of the inner disk radius and $\mu^{4/7}(GM)^{-1/7}\Dot{M}^{-2/7}$ \citep[][]{Pringle1972,Lamb1973}, where $\mu=BR^{\rm 3}$ is the magnetic moment. 

At the peak of the 2020 giant outburst, the bolometric flux and the corresponding QPO centroid frequency of 1A 0535+262 are $1.98\times10^{-7}~\rm{erg~cm^{-2}~s^{-1}}$ and 95\,mHz, respectively. Thus, for the BF model ($\nu_{\rm k}=\nu_{\rm q}+\nu_{\rm s}$) and KF model ($\nu_{\rm k}=\nu_{\rm q}$), $\aleph_{\rm{F}}$ can be given as $3.8\times10^{-5}~\rm{erg~cm^{-2}~s^{4/3}}$ and $4.8\times10^{-5}~\rm{erg~cm^{-2}~s^{4/3}}$, respectively. If we consider $B = 4.7\times 10^{12}$ G from the energy of the cyclotron line during the outburst peak \citep{Kong2021}, $\beta=1$ assumed by \citet{Camero2012}, $d=2.13$ kpc, $R=10^{6}$ cm and $M=2.8\times 10^{33}$\,g, $K$ would be $\sim 2.3$ and $\sim 2.4$, respectively. The values are similar to the result $\sim 2.4$ reported by \citet{Camero2012}. That is, the Alfven radius, $r_{\rm a}$, is about half of the inner radius of the Keplerian accretion disk, $r_{\rm k}$.

Considering that both the BF and KF models assume that the time scale of the initial variability generated at the inner radius of the accretion disk ($t_{\rm d}$) is the Keplerian time scale ($t_{\rm k}$), both models give that $r_{\rm k}$ is larger than $r_{\rm a}$. We suggest that $r_{\rm k}>r_{\rm a}$ because $t_{\rm d}$ and $t_{\rm k}$ may be not strictly equal, i.e. $t_{\rm d}=k_{\rm d}t_{\rm k}$, where $k_{\rm d}>1$ \citep{Tout1992, Stone1996}. In other words, the initial variability frequency is smaller than the Keplerian frequency, and the radius given by the initial variability frequency is larger than that from the Keplerian frequency. This explanation is consistent with the results given by \citet{Mushtukov2019} that in their simulations the Keplerian frequency is several times higher than the initial variability frequency.

During the full outburst, the relationship between the centroid frequency of the QPO and the bolometric flux is shown in Figure~\ref{fig:QPO_flux}. Following \citet{Finger1996}, the predicted power-law relation $F \propto \nu_{\rm k}^{7/3}$, given by Equation~(\ref{equ1}), was used to fit the data. Different from the 1994 outburst reported in \citet{Finger1996}, there is a clear discrepancy between the data and the theoretical relationship, both for the BF (black dashed line) and KF (black dotted line) models. Considering that the maximum frequency of the QPO in previous outbursts is around 70\,mHz, we fit again the data with a broken power-law by setting the break frequency at 70\,mHz (red lines in Figure~\ref{fig:QPO_flux}). The best-fitting power-law index ($\Gamma$) for the BF/KF model is $2.04\pm{0.16}/1.65\pm{0.09}$ (below 70\,mHz) and $1.84\pm{0.09}/1.63\pm{0.11}$ (above 70\,mHz), respectively. Nevertheless, when the centroid frequency of QPO is greater than 70\,mHz, a double-peak structure appears in the PDS, but in Figure~\ref{fig:QPO_flux}, the frequency during this period is given by fitting a single Lorentzian. We find a similar result if we use either frequency of the double peaks instead of the frequency given by the single Lorentzian and fit the relationship again. We note that the standard $F \propto \nu_{\rm k}^{7/3}$ relation from Equation~(\ref{equ1}), is established assuming that ${\aleph}_{F}$ is a constant. If the beaming factor $\beta$ in ${\aleph}_{F}$ is related to $\nu_{\rm k}$ or, more generally, to $F$, deviations from the standard relation would be expected.

\subsubsection{Energy-dependent properties of the QPO}
\label{dis:QPO_energy}

As discussed above, it can be seen that the relationship between the flux and the QPO frequency, either at the peak of the outburst or during the rest of the outburst, is not inconsistent with the BF and KF models in principle. In the following paragraphs, we will review the energy-dependent properties of the QPO.

Even though the count rate at the low energies is high enough, different from high energies (see Section~\ref{sec:QPO_energy}), the QPO is not significant below 27\,keV (see Figure~\ref{fig:PDS}). The emission at low energies is dominated by the accretion column and the thermal mound \citep{Kong2021}. If we consider the explanation of the BF model, the QPO should originate from the modulation of accretion rate from the disk. The variability propagates with the accreting material onto the polar caps of the pulsar, where most of X-rays are emitted. Therefore, the modulation should be imprinted into the emission from this region. The disappearance of the QPO signal at low energies may pose a challenge for the BF model.

For the KF model, the QPO is generated when the radiation is partially occulted by the inner edge of the disk. Therefore, the disappearance of the QPO at low energies may be due to a complete obscuration of the soft X-ray emitting region, as suggested by \citet{Camero2012}, or because the X-ray emitting region is not being obscured by the disk at all. However, the former case is directly contradicted by the fact that the energy spectra show strong soft X-ray emission \citep{Kong2021}. The latter case would require that the geometries of the disk are odd enough, to make sure that the high-energy emission region (the accretion column) is partially obscured but the low-energy region (the thermal mound and the accretion column) is not covered at all.

\begin{figure}
\centering
\includegraphics[width=\columnwidth]{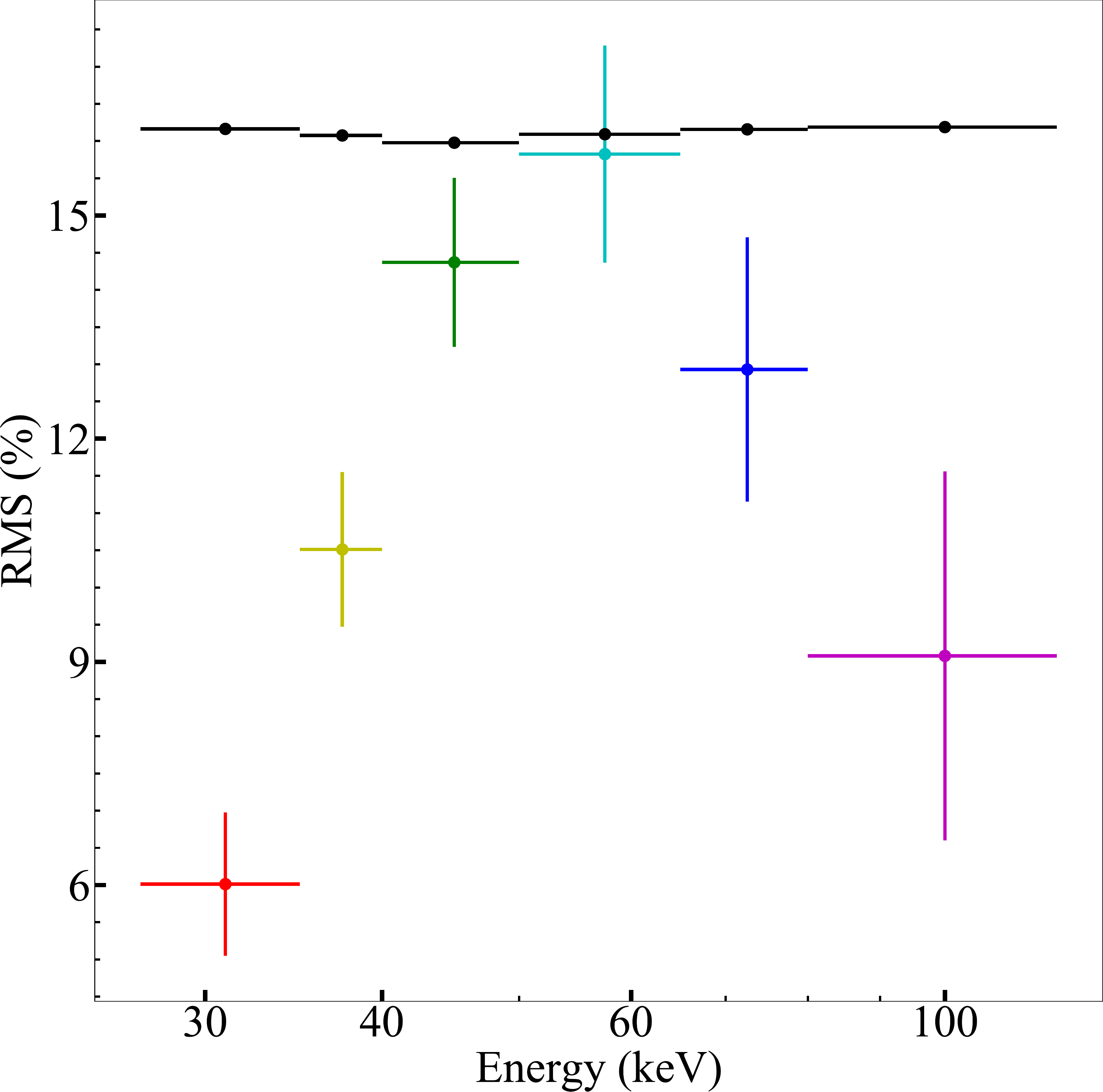} \\ %
\caption{Rms amplitude of the mHz QPO in 1A 0535+262 as a function of energy. The black points are the expected rms amplitude based on the BF model, and the colored points are the observed rms, the same as in Figure~\ref{fig:QPO_energy_evolution} .
\label{fig:Sim_RMS}}
\end{figure}

The fractional rms of the QPO, as shown in Figure~\ref{fig:QPO_energy_evolution}, is positively correlated with energy ($E$) when $E \lesssim 60$ keV. At higher energies, this trend is saturated and even appears to be negative, despite the large error bars. For the BF model, with the variability of the accreting materials propagating to the polar caps, we need to consider whether this model can produce the rms$-$energy relationship via interactions in the accretion column. The {\tt bwcycl} model in {\sc xspec} is used to do a simple simulation, as this model, taking the accretion rate into account, can account for the spectral formation process in accretion-powered X-ray pulsars based on a detailed treatment of the bulk and thermal Comptonization occurring in the accreting, shocked gas \citep[][]{Becker2007}. By assuming that the accretion rate at the outburst peak changing sinusoidally with a period 1/$v_{\rm q}\sim$10.5\,s, and given the following parameters from the spectral fitting: $\xi$=20 (parameter linked to the photon escape time), $\delta$=0.5 (ratio between bulk and thermal Comptonization importances), $T_{\rm e}$=8\,keV (electron temperature) and $r_{0}$=1200\,m (accretion column radius), we find that the rms does not evolve with energy when the accretion rate varies (as shown in Figure~\ref{fig:Sim_RMS}). That is, the variability from the accretion rate will not cause the observed rms$-$energy relationship. For the KF model the situation is a bit simpler. As the QPO is generated when the part of the radiation region is blocked by the accretion disk, it is difficult to understand the dependence of rms on energy as a result of obscuration.

\subsection{Other scenarios}
\label{sec:New model}

\subsubsection{Two jets in opposite directions}

From what has been discussed above, the origin of the QPO may not be well explained by the BF and KF models. Moreover, we further note that during the outburst peak there is a double-peak structure around the QPO frequency in the PDS, and the frequency difference between the two peaks remains more or less constant at $\sim$ 0.02\,Hz (about twice the spin frequency). This resembles the double-peak structure of kHz QPO in, e.g., Sco X-1 \citep{vdKlis1996}. One of the explanations is that the double-peak structure of kHz QPO originates from the emission reflected off inhomogeneities in two relativistic jets pointing in opposite directions near the NS, rather than directly from the NS \citep{vdKlis1997a,vdKlis1997b}. Indeed, radio emission, suggested as the jet emission, was recently observed in some strongly magnetized accreting X-ray pulsars, such as Swift J0243.6+6124 \citep{vdEijnden2018} and 1A 0535+262 \citep{vdEijnden2020}. 

Considering the Doppler effect, the QPO frequencies from the two relativistic jets would be shifted to $\nu_{\mp}=\nu_{\rm s}(1-v/c)/(1 \pm (v/c)\cos{\theta})$, where $v$ is the jet speed, and $\theta$ is the angle between the jets and the line of sight \citep{vdKlis1997a,vdKlis1997b}. However, if the QPO in 1A 0535+262 comes from two jets in opposite directions, from the frequency difference between the two peaks one jet would have to travel faster than the speed of light, and the other would have a negative speed. In addition, although the radio emission was detected in 1A 0535+262 \citep{vdEijnden2020}, the flux is about an order of magnitude lower than the jet emission of NS-LMXB and about two orders or magnitude lower than BH-LMXB, which makes the jet scenario unlikely.

\subsubsection{Precession of the neutron star}

The recent polarization results of the X-ray pulsar Her X-1 suggest that there is a misalignment (at least 20$^{\circ}$) between the NS spin axis and the binary orbital angular momentum \citep{Doroshenko2022}. If the NS spin axis of 1A 0535+262 is also inclined with respect to the orbital spin, we would expect that this system would precess around the spin axis and may produce the QPO signal in the accretion process, similar to the generation of low-frequency QPOs in some black-hole systems \citep[e.g.,][]{Ingram2011, Ingram2016, Huang2018, Liu2021}. In this picture, the generation of QPO's two-peak structure would be the result of beat frequency, and the difference between the two peaks would be the spin frequency of the NS or twice that. Considering that the QPO frequency in 1A 0535+262 is proportional to the source flux (Figures~\ref{fig:DyPDS} and \ref{fig:QPO_flux}), we estimate the precession frequency based on the conservation of angular momentum of accreting material. Assuming a misalignment angle $\phi$, and taking a spin-up rate of $\sim1.2\times10^{-11}$~Hz~s$^{-1}$ \citep{Finger1996, Camero2012}, and the source peak luminosity of $\sim 1\times10^{38}~\rm{erg~s^{-1}}$, the upper limit of the NS precession frequency is $\sim7\times10^{-4}\,\cos \phi$\,Hz, which is at least two orders of magnitude lower than the QPO frequency ($\sim 10^{-2}$\,Hz) observed in 1A 0535+262. This means that the NS precession resulting from the angular momentum transfer of accreting material, can not account for the QPO in this source.

\subsection{Implications}

Although there is currently no suitable model that fully explains the origin of the mHz QPO observed during the 2020 giant outburst of 1A 0535+262, here we report a series of interesting phenomena for the first time that should be accounted for in any model that aims at explaining this phenomena. First, the QPOs in 1A 0535+262 are detected above 80\,keV but not at the softest energy bands, below 27\,keV, which indicates that the QPOs must originate in the non-thermal component of the accretion column, and that the contribution of the thermal component to the QPO is very small. Second, the rms amplitude of the QPO first increases and then decreases with energy, reaching a maximum at 50--65\,keV, indicating that the amplitude of the variation of the non-thermal component (the accretion column) in the 50--65\,keV band is most dramatic. Finally, the QPO shows a double-peak structure during the peak of the outburst, implying that the structure of the accretion column or the physics of the radiation mechanism that produces the QPO changes at high luminosity. Some of the above results are analogous to what is observed in low-frequency QPOs in black-hole binaries, which have been proposed to be due to a resonance of the X-ray corona and the accretion disc \citep{Mastichiadis2022}. Considering some of the similarities between the mHz QPO in 1A 0535+262 and the low-frequency QPOs in black-hole binaries, the QPO in 1A 0535+262 could be due to a resonance between the accretion column and the accretion disc or the neutron-star surface (or both). Perhaps a model similar to the time-dependent Comptonisation model \citep{Karpouzas2020,Bellavita2022} that was used successfully to explain properties of low- and high-frequency QPOs in low-mass X-ray binaries \citep[e.g., ][]{Garcia2021,Mendez2022} needs to be developed for QPOs observed in systems like 1A 0535+262. For the moment, the origin of the QPOs in this source remains an open question, offering an opportunity for further theoretical research work.

\section{Summary}
\label{sec:con}

The high-statistic and broadband observations with \textit{Insight}-HXMT of the 2020 November giant outburst of 1A 0535+262 allow us to study the mHz QPOs in the source in great details. Our main findings are:\par

\begin{itemize}
\item[]
1. We have detected the mHz QPO in a broad energy band (27$-$120\,keV), and at the highest energy band so far for a QPO in among other similar sources (above 80\,keV). The QPO is most significant in the 50$-$65\,keV energy band. No significant QPO was detected below 27\,keV, which is consistent with previous reports. 
\end{itemize}

\begin{itemize}
\item[]
2. As the energy increases from 27\,keV to 120\,keV, the fractional rms of the QPO first increases and then decreases, peaking at 50$-$65\,keV. 
\end{itemize}

\begin{itemize}
\item[]
3. The centroid frequency of the QPO is positively correlated to the bolometric flux. Beaming effects may cause this relationship to slightly deviate from the standard relationship.
\end{itemize}

\begin{itemize}
\item[]
4. Neither the KF nor the BF models can explain the energy-dependent behavior of the QPO well, including the rms-energy relationship and the non-detection of the QPO at low energies. 
\end{itemize}

\begin{itemize}
\item[]
5. A double-peak structure of the QPO is observed at the peak of the outburst, with the frequency difference between the two peaks being constant at twice the spin frequency of the NS. Neither a model of two jets in opposite directions nor NS precession may explain the origin of the QPO. 
\end{itemize}

\begin{itemize}
\item[]
6. By considering that the QPO is detected above 27\,keV, especially the detection above 80\,keV, we conclude that the QPO originates from the non-thermal component (e.g., the interaction of the accretion column with the disc or the NS surface or both). Further theoretical work is necessary to understand the origin and energy-dependent behavior of the QPO.
\end{itemize}

\section*{Acknowledgements}

We thank the anonymous referee and Yuexin Zhang for useful comments that have improved the paper. This work made use of data from the Insight-HXMT mission, a project funded by China National Space Administration (CNSA) and the Chinese Academy of Sciences (CAS). This work is supported by the National Key R\&D Program of China (2021YFA0718500). We acknowledges funding support from the National Natural Science Foundation of China (NSFC) under grant Nos. 12122306, U1838115, U1838201, U1838202, 12173103, U2038101, U1938103, 11733009, the CAS Pioneer Hundred Talent Program Y8291130K2 and the Scientific and technological innovation project of IHEP Y7515570U1. R.M. acknowledges support from China Scholarship Council (CSC 202104910402). M.M. acknowledges the research programme Athena with project number 184.034.002, which is (partly) financed by the Dutch Research Council (NWO).

{\it Facilities:} \textit{Insight-HXMT}


\section*{Data Availability}

The data for \textit{Insight-HXMT} underlying this article is available format at \textit{Insight-HXMT} website (\href{http://archive.hxmt.cn/proposal}{http://archive.hxmt.cn/proposal}; data in compressed format).







\appendix

\section{Tables}


\begin{center}
\onecolumn
\begin{longtable}{cccccc}
\caption{\textit{Insight}-HXMT observations log, and the net count rates in the HE band (27$-$120\,keV).}
\label{tab:obs_info}\\
\hline
ObsID & ExpID & Start Time & Start Time & HE Exp.~Time & HE Count Rate \\
 &   & (day) & (MJD) & (s) & (cts $\rm{s^{-1}}$) \\
\hline
P0314316001$^{*}$ & 01 & 2020-11-14 07:49:43 & 59167.33 & 593.8 & $2450\pm{4}$ \\
  & 02 & 2020-11-14 11:42:54 & 59167.49 & 4328.1 & $2363.4\pm{1.4}$ \\
  & 03 & 2020-11-14 15:05:07 & 59167.63 & 1990.5 & $2532\pm{2}$ \\
  & 05 & 2020-11-14 21:26:33 & 59167.89 & 1427.3 & $2962\pm{2}$ \\
  & 06 & 2020-11-15 00:37:20 & 59168.03 & 2612.7 & $2713.1\pm{1.8}$ \\
  & 07 & 2020-11-15 03:48:06 & 59168.16 & 1148.8 & $2956\pm{3}$ \\
  & 08 & 2020-11-15 06:59:01 & 59168.29 & 721.4 & $3037\pm{3}$ \\
  & 09 & 2020-11-15 10:09:48 & 59168.42 & 2242.2 & $3102\pm{2}$ \\
  & 10 & 2020-11-15 13:19:10 & 59168.56 & 4115.3 & $3138.0\pm{1.5}$ \\
  & 12 & 2020-11-15 19:42:08 & 59168.82 & 2867.6 & $3325.4\pm{1.8}$ \\
  & 13 & 2020-11-15 22:52:47 & 59168.95 & 1427.4 & $3482\pm{2}$ \\
\hline  
P0314316002$^{*}$ & 01 & 2020-11-16 05:56:48 & 59169.3 & 598.1 & $3310\pm{4}$ \\
  & 02 & 2020-11-16 09:41:45 & 59169.40 & 2664.3 & $3540.6\pm{1.9}$ \\
  & 03 & 2020-11-16 13:09:58 & 59169.55 & 3768.2 & $3565.5\pm{1.6}$ \\
  & 04 & 2020-11-16 16:22:12 & 59169.68 & 343.1 & $3636.1\pm{5.1}$ \\
  & 05 & 2020-11-16 19:32:59 & 59169.82 & 2822.0 & $3747.2\pm{1.8}$ \\
  & 06 & 2020-11-16 22:43:39 & 59169.95 & 2760.1 & $3813.8\pm{1.8}$ \\
  & 07 & 2020-11-17 01:54:26 & 59170.08 & 1809.3 & $3706\pm{2}$ \\
  & 08 & 2020-11-17 05:05:19 & 59170.21 & 539.8 & $3815\pm{4}$ \\
  & 09 & 2020-11-17 08:16:06 & 59170.35 & 849.3 & $3884\pm{3}$ \\
  & 10 & 2020-11-17 11:18:06 & 59170.47 & 4313.3 & $3887.9\pm{1.5}$ \\
  & 11 & 2020-11-17 14:37:40 & 59170.61 & 1216.0 & $4116\pm{3}$ \\
  & 12 & 2020-11-17 17:48:27 & 59170.74 & 2077.0 & $3876\pm{2}$ \\
  & 13 & 2020-11-17 20:59:06 & 59170.88 & 2839.2 & $4053.6\pm{1.9}$ \\
  & 14 & 2020-11-18 00:09:52 & 59171.01 & 1322.2 & $4021\pm{3}$ \\
\hline  
P0314316003$^{*}$ & 01 & 2020-11-18 07:14:28 & 59171.3 & 1340.9 & $4177\pm{3}$ \\
  & 02 & 2020-11-18 11:09:23 & 59171.47 & 4227.0 & $4255.1\pm{1.5}$ \\
  & 03 & 2020-11-18 14:28:32 & 59171.60 & 389.0 & $4391\pm{5}$ \\
  & 04 & 2020-11-18 17:39:18 & 59171.74 & 2485.2 & $4038\pm{2}$ \\
  & 05 & 2020-11-18 20:49:57 & 59171.87 & 2827.4 & $4140.6\pm{1.8}$ \\
  & 06 & 2020-11-19 00:00:44 & 59172.00 & 2156.5 & $4138\pm{2}$ \\
  & 07 & 2020-11-19 03:11:31 & 59172.13 & 299.7 & $4022\pm{5}$ \\
  & 08 & 2020-11-19 06:22:27 & 59172.27 & 277.8 & $3953\pm{6}$ \\
  & 09 & 2020-11-19 09:33:14 & 59172.40 & 3615.3 & $4082.9\pm{1.7}$ \\
  & 10 & 2020-11-19 12:42:00 & 59172.53 & 1835.8 & $3990\pm{2}$ \\
  & 11 & 2020-11-19 15:54:48 & 59172.66 & 1321.9 & $4019\pm{3}$ \\
  & 12 & 2020-11-19 19:05:28 & 59172.80 & 2857.0 & $4024.0\pm{1.8}$ \\
  & 13 & 2020-11-19 22:16:23 & 59172.93 & 560.8 & $4356\pm{4}$ \\
  & 14 & 2020-11-20 01:27:03 & 59173.06 & 925.3 & $4325\pm{3}$ \\
  & 15 & 2020-11-20 04:37:57 & 59173.19 & 421.6 & $4322\pm{5}$ \\
  & 16 & 2020-11-20 07:48:45 & 59173.33 & 1625.4 & $4160\pm{3}$ \\
  & 17 & 2020-11-20 10:51:33 & 59173.45 & 3846.9 & $4101.6\pm{1.6}$ \\
  & 19 & 2020-11-20 17:21:07 & 59173.72 & 2808.3 & $4146.0\pm{1.9}$ \\
  & 20 & 2020-11-20 20:31:47 & 59173.86 & 2801.2 & $4091.5\pm{1.8}$ \\
\hline  
P0314316004$^{*}$ & 01 & 2020-11-21 05:12:10 & 59174.2 & 554.5 & $4250\pm{4}$ \\
  & 02 & 2020-11-21 08:57:49 & 59174.37 & 3773.1 & $3968.8\pm{1.6}$ \\
  & 03 & 2020-11-21 12:23:09 & 59174.52 & 1706.4 & $3992\pm{2}$ \\
  & 04 & 2020-11-21 15:36:42 & 59174.65 & 1398.4 & $3909\pm{3}$ \\
  & 05 & 2020-11-21 18:47:22 & 59174.78 & 2853.7 & $3881.5\pm{1.8}$ \\
  & 06 & 2020-11-21 21:58:10 & 59174.92 & 2574.6 & $3943.5\pm{1.9}$ \\
  & 07 & 2020-11-22 01:08:58 & 59175.05 & 805.6 & $4136\pm{3}$ \\
  & 08 & 2020-11-22 04:19:54 & 59175.18 & 294.2 & $3771\pm{5}$ \\
  & 09 & 2020-11-22 07:30:42 & 59175.31 & 2325.4 & $3945\pm{2}$ \\
  & 10 & 2020-11-22 10:33:15 & 59175.44 & 3554.2 & $3964.6\pm{1.7}$ \\
  & 11 & 2020-11-22 13:52:19 & 59175.58 & 252.6 & $4022\pm{6}$ \\
  & 12 & 2020-11-22 17:03:08 & 59175.71 & 2824.4 & $3974.6\pm{1.9}$ \\
  & 13 & 2020-11-22 20:13:48 & 59175.84 & 2768.6 & $3956.0\pm{1.8}$ \\
  & 14 & 2020-11-22 23:24:45 & 59175.98 & 1813.0 & $3821\pm{2}$ \\
\hline  
P0314316005 & 04 & 2020-11-24 16:45:29 & 59177.7 & 2854.8 & $3740.8\pm{1.8}$ \\
  & 05 & 2020-11-24 19:56:11 & 59177.83 & 2722.7 & $3689.1\pm{1.8}$ \\
  & 06 & 2020-11-24 23:07:10 & 59177.96 & 1034.3 & $3682\pm{3}$ \\
  & 07 & 2020-11-25 02:17:54 & 59178.10 & 373.0 & $3526\pm{5}$ \\
  & 08 & 2020-11-25 05:28:52 & 59178.23 & 1350.1 & $3551\pm{3}$ \\
  & 09 & 2020-11-25 08:39:43 & 59178.36 & 3876.7 & $3623.6\pm{1.6}$ \\
  & 11 & 2020-11-25 15:01:25 & 59178.63 & 2723.6 & $3768.2\pm{1.9}$ \\
  & 12 & 2020-11-25 18:12:09 & 59178.76 & 2817.4 & $3578.2\pm{1.8}$ \\
  & 13 & 2020-11-25 21:23:09 & 59178.89 & 2102.5 & $3566\pm{2}$ \\
  & 14 & 2020-11-26 00:33:53 & 59179.02 & 250.3 & $3847\pm{6}$ \\
  & 15 & 2020-11-26 03:44:53 & 59179.16 & 323.8 & $3641\pm{5}$ \\
  & 16 & 2020-11-26 06:55:45 & 59179.29 & 3630.4 & $3642.8\pm{1.6}$ \\
  & 17 & 2020-11-26 09:55:52 & 59179.41 & 1756.1 & $3435\pm{2}$ \\
  & 18 & 2020-11-26 13:15:38 & 59179.55 & 1368.1 & $3619\pm{3}$ \\
  & 19 & 2020-11-26 16:28:16 & 59179.69 & 2856.2 & $3561.15\pm{1.8}$ \\
  & 20 & 2020-11-26 19:39:10 & 59179.82 & 2642.6 & $3353.0\pm{1.8}$ \\
  & 21 & 2020-11-26 22:50:11 & 59179.95 & 747.8 & $3201\pm{3}$ \\
\hline  
P0314316006 & 01 & 2020-11-27 05:54:16 & 59180.3 & 3811.8 & $3482.1\pm{1.6}$ \\
  & 02 & 2020-11-27 09:46:25 & 59180.41 & 1726.0 & $3349\pm{2}$ \\
  & 03 & 2020-11-27 13:06:11 & 59180.55 & 1400.0 & $3306.3\pm{2.6}$ \\
  & 04 & 2020-11-27 16:20:03 & 59180.68 & 2854.4 & $3477.2\pm{1.8}$ \\
  & 05 & 2020-11-27 19:31:07 & 59180.81 & 1364.2 & $3321\pm{3}$ \\
  & 06 & 2020-11-27 22:42:03 & 59180.95 & 126.2 & $3049\pm{8}$ \\
  & 07 & 2020-11-28 01:52:52 & 59181.08 & 354.2 & $3134\pm{5}$ \\
  & 08 & 2020-11-28 05:03:56 & 59181.21 & 2166.0 & $3232\pm{2}$ \\
  & 09 & 2020-11-28 08:14:53 & 59181.34 & 3604.5 & $3255.3\pm{1.6}$ \\
  & 10 & 2020-11-28 11:16:46 & 59181.47 & 260.5 & $3204\pm{6}$ \\
  & 11 & 2020-11-28 14:36:48 & 59181.61 & 2863.2 & $3116.9\pm{1.8}$ \\
  & 12 & 2020-11-28 17:47:38 & 59181.74 & 2777.5 & $3131.2\pm{1.8}$ \\
  & 13 & 2020-11-28 20:58:45 & 59181.87 & 1840.1 & $3206\pm{2}$ \\
  & 14 & 2020-11-29 00:09:36 & 59182.01 & 229.0 & $2807\pm{6}$ \\
  & 15 & 2020-11-29 03:20:36 & 59182.14 & 698.6 & $3146\pm{4}$ \\
  & 16 & 2020-11-29 06:31:44 & 59182.27 & 3948.2 & $3143.7\pm{1.5}$ \\
  & 17 & 2020-11-29 09:42:44 & 59182.41 & 1427.2 & $3065\pm{3}$ \\
  & 18 & 2020-11-29 12:47:22 & 59182.53 & 2253.1 & $2948\pm{2}$ \\
  & 19 & 2020-11-29 16:04:39 & 59182.67 & 2890.7 & $3012.2\pm{1.7}$ \\
  & 20 & 2020-11-29 19:15:49 & 59182.80 & 2271.6 & $2815.9\pm{1.9}$ \\
  & 21 & 2020-11-29 22:26:52 & 59182.94 & 319.5 & $3552\pm{5}$ \\
\hline  
P0314316008 & 01 & 2020-11-30 07:04:36 & 59183.3 & 4387.3 & $2801.2\pm{1.4}$ \\
  & 02 & 2020-11-30 10:57:47 & 59183.46 & 1130.8 & $2873\pm{3}$ \\
  & 03 & 2020-11-30 14:19:09 & 59183.60 & 3267.7 & $2797.9\pm{1.7}$ \\
  & 04 & 2020-11-30 17:33:16 & 59183.73 & 2935.8 & $2631.8\pm{1.7}$ \\
  & 05 & 2020-11-30 20:44:31 & 59183.87 & 1185.5 & $2692\pm{3}$ \\
  & 06 & 2020-11-30 23:55:39 & 59184.00 & 353.8 & $2775\pm{5}$ \\
  & 07 & 2020-12-01 03:06:40 & 59184.13 & 1271.4 & $2853\pm{3}$ \\
  & 08 & 2020-12-01 06:17:59 & 59184.26 & 3912.6 & $2595.2\pm{1.5}$ \\
  & 09 & 2020-12-01 09:29:10 & 59184.40 & 18.4 & $3094\pm{20}$ \\
  & 10 & 2020-12-01 12:40:22 & 59184.53 & 3562.8 & $2421\pm{1.6}$ \\
  & 11 & 2020-12-01 15:51:28 & 59184.66 & 3439.3 & $2458\pm{1.6}$ \\
  & 12 & 2020-12-01 19:02:50 & 59184.79 & 2595.2 & $2621.8\pm{1.8}$ \\
  & 13 & 2020-12-01 22:14:06 & 59184.93 & 816.3 & $2406\pm{3}$ \\
  & 14 & 2020-12-02 02:11:32 & 59185.09 & 1523.9 & $2584\pm{2}$ \\
  & 15 & 2020-12-02 05:37:47 & 59185.24 & 2063.8 & $2514.1\pm{2.0}$ \\
  & 16 & 2020-12-02 07:36:22 & 59185.32 & 1762.5 & $2448\pm{2}$ \\
  & 17 & 2020-12-02 10:59:22 & 59185.46 & 2063.3 & $2444\pm{2}$ \\
  & 18 & 2020-12-02 14:10:45 & 59185.59 & 3931.1 & $2520.3\pm{1.5}$ \\
  & 19 & 2020-12-02 17:22:10 & 59185.72 & 3545.1 & $2429.5\pm{1.5}$ \\
  & 20 & 2020-12-02 20:33:38 & 59185.86 & 1852.1 & $2326\pm{2}$ \\
\hline  
P0314316009 & 01 & 2020-12-03 05:06:55 & 59186.2 & 4866.5 & $2421.1\pm{1.3}$ \\
  & 02 & 2020-12-03 08:49:28 & 59186.37 & 82.8 & $2384\pm{11}$ \\
  & 03 & 2020-12-03 12:10:02 & 59186.51 & 11228.6 & $2285.9\pm{0.9}$ \\
  & 04 & 2020-12-03 21:09:49 & 59186.88 & 1524.9 & $2150\pm{2}$ \\
  & 05 & 2020-12-04 00:14:19 & 59187.01 & 541.5 & $2329\pm{4}$ \\
  & 06 & 2020-12-04 03:35:23 & 59187.15 & 4295.4 & $2232.5\pm{1.4}$ \\
  & 07 & 2020-12-04 07:00:09 & 59187.29 & 1553.9 & $2173\pm{2}$ \\
  & 08 & 2020-12-04 10:19:58 & 59187.43 & 8147.7 & $2117.8\pm{1.0}$ \\
  & 09 & 2020-12-04 20:59:03 & 59187.88 & 1987.5 & $2185.7\pm{1.9}$ \\
  & 10 & 2020-12-05 00:04:24 & 59188.00 & 659.1 & $2032\pm{4}$ \\
  & 11 & 2020-12-05 03:27:02 & 59188.14 & 2937.0 & $2045.5\pm{1.7}$ \\
  & 12 & 2020-12-05 06:50:39 & 59188.29 & 163.6 & $2080\pm{7}$ \\
  & 13 & 2020-12-05 10:10:33 & 59188.42 & 7478.8 & $1873.0\pm{1.1}$ \\
  & 14 & 2020-12-05 19:18:23 & 59188.81 & 4089.5 & $1885.0\pm{1.4}$ \\
\hline  
P0314316010 & 01 & 2020-12-06 06:25:59 & 59189.3 & 1175.4 & $1775\pm{3}$ \\
  & 02 & 2020-12-06 10:01:09 & 59189.42 & 10656.1 & $1713.9\pm{0.9}$ \\
  & 03 & 2020-12-06 19:05:38 & 59189.80 & 2731.4 & $1877.6\pm{1.6}$ \\
  & 04 & 2020-12-06 22:07:25 & 59189.92 & 1267.6 & $1558\pm{3}$ \\
  & 05 & 2020-12-07 01:22:44 & 59190.06 & 3755.1 & $1739.4\pm{1.5}$ \\
  & 06 & 2020-12-07 04:51:35 & 59190.20 & 2217.0 & $1678.0\pm{1.8}$ \\
  & 07 & 2020-12-07 08:11:27 & 59190.34 & 10217.1 & $1630.5\pm{0.9}$ \\
  & 08 & 2020-12-07 18:54:53 & 59190.79 & 2786.5 & $1676.9\pm{1.5}$ \\
  & 09 & 2020-12-07 21:57:28 & 59190.92 & 718.4 & $1696\pm{3}$ \\
  & 10 & 2020-12-08 01:14:04 & 59191.05 & 3790.2 & $1572.8\pm{1.5}$ \\
  & 11 & 2020-12-08 04:42:11 & 59191.20 & 1794.7 & $1460.5\pm{2.0}$ \\
  & 12 & 2020-12-08 08:01:59 & 59191.34 & 2141.5 & $1571.3\pm{1.9}$ \\
  & 13 & 2020-12-08 11:57:19 & 59191.50 & 3333.2 & $1416.7\pm{1.5}$ \\
  & 14 & 2020-12-08 15:07:34 & 59191.63 & 3506.5 & $1351.0\pm{1.4}$ \\
  & 15 & 2020-12-08 18:17:49 & 59191.76 & 2924.8 & $1394.3\pm{1.5}$ \\
  & 16 & 2020-12-08 21:28:15 & 59191.90 & 258.9 & $1253\pm{5}$ \\
\hline  
P0314316011 & 01 & 2020-12-09 04:31:43 & 59192.2 & 1400.9 & $1413\pm{2}$ \\
  & 02 & 2020-12-09 07:52:31 & 59192.33 & 2264.2 & $1239.8\pm{1.8}$ \\
  & 03 & 2020-12-09 11:45:49 & 59192.49 & 2938.1 & $1258.9\pm{1.5}$ \\
  & 04 & 2020-12-09 14:56:25 & 59192.62 & 3103.6 & $1298.4\pm{1.5}$ \\
  & 05 & 2020-12-09 18:06:55 & 59192.76 & 2845.1 & $1158.0\pm{1.5}$ \\
  & 07 & 2020-12-10 00:28:16 & 59193.02 & 2789.0 & $1200.7\pm{1.6}$ \\
  & 08 & 2020-12-10 03:38:54 & 59193.15 & 1454.1 & $1285\pm{2}$ \\
  & 09 & 2020-12-10 06:49:33 & 59193.29 & 681.6 & $1160\pm{3}$ \\
  & 10 & 2020-12-10 10:00:11 & 59193.42 & 1962.7 & $1352.2\pm{1.8}$ \\
  & 11 & 2020-12-10 13:10:51 & 59193.55 & 2718.3 & $1243.8\pm{1.6}$ \\
  & 12 & 2020-12-10 16:21:22 & 59193.68 & 2766.7 & $1129.2\pm{1.5}$ \\
  & 13 & 2020-12-10 19:32:10 & 59193.81 & 1489.2 & $1079\pm{2}$ \\
  & 14 & 2020-12-10 22:42:41 & 59193.95 & 1184.9 & $1109\pm{2}$ \\
  & 15 & 2020-12-11 01:53:30 & 59194.08 & 2181.4 & $1164.3\pm{1.8}$ \\
  & 16 & 2020-12-11 05:04:10 & 59194.21 & 172.9 & $1067\pm{6}$ \\
  & 17 & 2020-12-11 08:14:50 & 59194.34 & 1812.0 & $1117.0\pm{2.0}$ \\
  & 18 & 2020-12-11 11:25:31 & 59194.48 & 2344.5 & $1052.3\pm{1.6}$ \\
  & 19 & 2020-12-11 14:36:03 & 59194.61 & 2585.0 & $1014.5\pm{1.5}$ \\
  & 20 & 2020-12-11 17:46:44 & 59194.74 & 2861.9 & $1017.2\pm{1.5}$ \\
  & 22 & 2020-12-12 00:08:14 & 59195.01 & 1551.0 & $1023\pm{2}$ \\
\hline  
P0314316012 & 01 & 2020-12-12 07:15:06 & 59195.3 & 3036.7 & $881.3\pm{1.7}$ \\
  & 02 & 2020-12-12 11:15:32 & 59195.47 & 2364.8 & $933.2\pm{1.6}$ \\
  & 03 & 2020-12-12 14:26:12 & 59195.60 & 2606.3 & $892.9\pm{1.5}$ \\
  & 04 & 2020-12-12 17:37:02 & 59195.73 & 2696.0 & $885.1\pm{1.6}$ \\
  & 05 & 2020-12-12 20:47:43 & 59195.87 & 243.4 & $787\pm{5}$ \\
  & 06 & 2020-12-12 23:58:17 & 59196.00 & 2150.7 & $884.3\pm{1.7}$ \\
  & 07 & 2020-12-13 03:09:07 & 59196.13 & 34.9 & $1237\pm{11}$ \\
  & 08 & 2020-12-13 06:19:49 & 59196.26 & 1425.0 & $872\pm{2}$ \\
  & 09 & 2020-12-13 09:30:23 & 59196.40 & 2247.0 & $885.5\pm{1.7}$ \\
  & 10 & 2020-12-13 12:41:05 & 59196.53 & 2507.0 & $787.8\pm{1.5}$ \\
  & 11 & 2020-12-13 15:51:55 & 59196.66 & 2784.7 & $761.4\pm{1.4}$ \\
  & 12 & 2020-12-13 19:02:37 & 59196.79 & 850.3 & $737\pm{3}$ \\
  & 13 & 2020-12-13 22:13:19 & 59196.93 & 1837.2 & $802.2\pm{2.0}$ \\
  & 14 & 2020-12-14 01:24:01 & 59197.06 & 1197.4 & $682\pm{2}$ \\
  & 15 & 2020-12-14 04:34:43 & 59197.19 & 296.3 & $935\pm{5}$ \\
  & 17 & 2020-12-14 10:56:00 & 59197.46 & 2407.7 & $739.6\pm{1.5}$ \\
  & 18 & 2020-12-14 14:06:42 & 59197.59 & 2651.6 & $615.5\pm{1.4}$ \\
  & 19 & 2020-12-14 17:17:33 & 59197.72 & 1835.9 & $659.4\pm{1.8}$ \\
  & 20 & 2020-12-14 20:28:15 & 59197.85 & 846.0 & $608\pm{3}$ \\
\hline  
P0314316013 & 01 & 2020-12-15 05:11:30 & 59198.2 & 1914.8 & $550.3\pm{1.9}$ \\
  & 02 & 2020-12-15 09:10:58 & 59198.38 & 2282.4 & $518.0\pm{1.6}$ \\
  & 03 & 2020-12-15 12:21:41 & 59198.52 & 2546.2 & $548.7\pm{1.4}$ \\
  & 04 & 2020-12-15 15:32:31 & 59198.65 & 2844.3 & $529.6\pm{1.4}$ \\
  & 05 & 2020-12-15 18:43:13 & 59198.78 & 23.4 & $443\pm{16}$ \\
  & 06 & 2020-12-15 21:53:56 & 59198.91 & 2216.9 & $537.0\pm{1.7}$ \\
  & 07 & 2020-12-16 01:04:39 & 59199.05 & 593.2 & $457\pm{3}$ \\
  & 08 & 2020-12-16 04:15:22 & 59199.18 & 682.1 & $609\pm{3}$ \\
  & 09 & 2020-12-16 07:26:04 & 59199.31 & 2345.2 & $480.5\pm{1.6}$ \\
  & 10 & 2020-12-16 10:36:39 & 59199.44 & 2453.5 & $476.3\pm{1.5}$ \\
  & 11 & 2020-12-16 13:47:23 & 59199.58 & 2700.3 & $400.6\pm{1.4}$ \\
  & 12 & 2020-12-16 16:58:13 & 59199.71 & 1486.9 & $472.0\pm{1.9}$ \\
  & 13 & 2020-12-16 20:08:56 & 59199.84 & 1149.0 & $504\pm{2}$ \\
  & 14 & 2020-12-16 23:19:32 & 59199.97 & 1139.9 & $424\pm{2}$ \\
  & 16 & 2020-12-17 05:41:06 & 59200.24 & 1832.9 & $415.8\pm{1.9}$ \\
  & 17 & 2020-12-17 08:51:41 & 59200.37 & 2342.6 & $337.6\pm{1.6}$ \\
  & 18 & 2020-12-17 12:02:24 & 59200.50 & 2585.1 & $392.6\pm{1.5}$ \\
  & 19 & 2020-12-17 15:13:15 & 59200.63 & 2897.6 & $386.1\pm{1.4}$ \\
  & 21 & 2020-12-17 21:34:42 & 59200.90 & 2317.1 & $365.4\pm{1.7}$ \\
\hline  
P0314316014 & 01 & 2020-12-18 06:18:40 & 59201.3 & 2930.0 & $354.7\pm{1.5}$ \\
  & 02 & 2020-12-18 10:17:27 & 59201.43 & 2492.4 & $331.2\pm{1.4}$ \\
  & 03 & 2020-12-18 13:28:19 & 59201.56 & 2753.9 & $293.5\pm{1.4}$ \\
  & 04 & 2020-12-18 16:39:02 & 59201.69 & 1329.9 & $277\pm{2}$ \\
  & 05 & 2020-12-18 19:49:45 & 59201.83 & 1347.9 & $293\pm{2}$ \\
  & 06 & 2020-12-18 23:00:22 & 59201.96 & 1017.4 & $352\pm{2}$ \\
  & 07 & 2020-12-19 02:11:13 & 59202.09 & 15.5 & $336\pm{17}$ \\
  & 08 & 2020-12-19 05:21:56 & 59202.22 & 1803.0 & $285.4\pm{1.8}$ \\
  & 09 & 2020-12-19 08:32:32 & 59202.36 & 2384.9 & $301.3\pm{1.5}$ \\
  & 10 & 2020-12-19 11:43:16 & 59202.49 & 2626.8 & $257.6\pm{1.4}$ \\
  & 11 & 2020-12-19 14:54:07 & 59202.62 & 2435.8 & $262.9\pm{1.5}$ \\
  & 12 & 2020-12-19 18:04:51 & 59202.75 & 535.1 & $260\pm{3}$ \\
  & 13 & 2020-12-19 21:15:35 & 59202.89 & 14650.1 & $207.5\pm{0.7}$ \\
\hline  
P0314316015 & 01 & 2020-12-21 05:50:46 & 59204.2 & 1874.6 & $159.6\pm{1.7}$ \\
  & 02 & 2020-12-21 09:48:53 & 59204.41 & 2543.8 & $129.5\pm{1.5}$ \\
  & 03 & 2020-12-21 12:59:45 & 59204.54 & 2839.5 & $138.3\pm{1.4}$ \\
  & 04 & 2020-12-21 16:10:29 & 59204.67 & 60.4 & $218\pm{8}$ \\
  & 05 & 2020-12-21 19:21:14 & 59204.81 & 2330.4 & $137.7\pm{1.6}$ \\
  & 06 & 2020-12-21 22:31:50 & 59204.94 & 825.1 & $116\pm{3}$ \\
  & 07 & 2020-12-22 01:42:42 & 59205.07 & 680.6 & $119\pm{3}$ \\
  & 08 & 2020-12-22 04:53:27 & 59205.20 & 2358.3 & $132.2\pm{1.6}$ \\
  & 09 & 2020-12-22 08:04:04 & 59205.34 & 2450.6 & $107.6\pm{1.5}$ \\
  & 10 & 2020-12-22 11:14:49 & 59205.47 & 2696.5 & $111.1\pm{1.4}$ \\
  & 11 & 2020-12-22 14:25:40 & 59205.60 & 1484.6 & $115.2\pm{1.9}$ \\
  & 12 & 2020-12-22 17:36:25 & 59205.73 & 1213.2 & $113\pm{2}$ \\
  & 13 & 2020-12-22 20:47:09 & 59205.87 & 1814.6 & $94.9\pm{1.9}$ \\
  & 15 & 2020-12-23 03:08:39 & 59206.13 & 1932.9 & $105.4\pm{1.8}$ \\
  & 17 & 2020-12-23 09:30:00 & 59206.40 & 2581.9 & $88.6\pm{1.4}$ \\
  & 18 & 2020-12-23 12:40:53 & 59206.53 & 2899.2 & $99.9\pm{1.3}$ \\
  & 20 & 2020-12-23 19:02:23 & 59206.79 & 2768.8 & $87.2\pm{1.5}$ \\
  & 21 & 2020-12-23 22:13:00 & 59206.93 & 125.1 & $92.8\pm{7}$ \\
  & 22 & 2020-12-24 01:23:53 & 59207.06 & 744.5 & $74\pm{3}$ \\
  & 23 & 2020-12-24 04:34:28 & 59207.19 & 2460.1 & $75.8\pm{1.5}$ \\
  & 24 & 2020-12-24 07:45:15 & 59207.32 & 2521.1 & $69.0\pm{1.4}$ \\
  & 25 & 2020-12-24 10:56:08 & 59207.46 & 2747.5 & $58.7\pm{1.4}$ \\
  & 26 & 2020-12-24 14:06:53 & 59207.59 & 1386.8 & $65\pm{2}$ \\
  & 27 & 2020-12-24 17:17:38 & 59207.72 & 1570.6 & $62.2\pm{1.9}$ \\
  & 28 & 2020-12-24 20:28:23 & 59207.85 & 2138.7 & $60.9\pm{1.8}$ \\
\hline   
\end{longtable}
\end{center}
\footnotesize{Notes: $^{*}$ indicates the ObsIDs with the  double-peak structure.}\newline

\bsp	
\label{lastpage}
\end{document}